\documentclass[review,number,sort&compress]{elsarticle}

\usepackage{lineno,hyperref}
\modulolinenumbers[5]

\journal{Journal of \LaTeX\ Templates}
\usepackage{lscape}

\begin{document}

\begin{frontmatter}

  \title{Development of low radioactivity photomultiplier tubes
    for the XMASS-I detector}

  \address{\rm\normalsize XMASS Collaboration$^*$}
  \cortext[cor1]{{\it E-mail address:} xmass.publications8@km.icrr.u-tokyo.ac.jp .} 
  
  \author[ICRR,IPMU]{K.~Abe}
  \author[ICRR,IPMU]{K.~Hiraide}
  \author[ICRR,IPMU]{K.~Ichimura}
  \author[ICRR,IPMU]{Y.~Kishimoto}
  \author[ICRR,IPMU]{K.~Kobayashi}
  \author[ICRR]{M.~Kobayashi}
  \author[ICRR,IPMU]{S.~Moriyama}
  \author[ICRR,IPMU]{M.~Nakahata}
  \author[ICRR]{T.~Norita}
  \author[ICRR,IPMU]{H.~Ogawa\fnref{Ogawanow}}
  \author[ICRR]{K.~Sato}
  \author[ICRR,IPMU]{H.~Sekiya}
  \author[ICRR]{O.~Takachio}
  \author[ICRR,IPMU]{A.~Takeda}
  \author[ICRR]{S.~Tasaka}
  \author[ICRR,IPMU]{M.~Yamashita}
  \author[ICRR,IPMU]{B.~S.~Yang\fnref{Yangnow}}
  \author[IBS]{N.~Y.~Kim}
  \author[IBS]{Y.~D.~Kim}
  \author[ISEE,KMI]{Y.~Itow}
  \author[ISEE]{K.~Kanzawa}
  \author[ISEE]{R.~Kegasa}
  \author[ISEE]{K.~Masuda}
  \author[ISEE]{H.~Takiya}
  \author[Tokushima]{K.~Fushimi\fnref{Tokushimanow}}
  \author[Tokushima]{G.~Kanzaki}
  \author[IPMU]{K.~Martens}
  \author[IPMU]{Y.~Suzuki}
  \author[IPMU]{B.~D.~Xu}
  \author[Kobe]{R.~Fujita}
  \author[Kobe]{K.~Hosokawa\fnref{Tohokunow}}
  \author[Kobe]{K.~Miuchi}
  \author[Kobe]{N.~Oka}
  \author[Kobe,IPMU]{Y.~Takeuchi}
  \author[KRISS,IBS]{Y.~H.~Kim}
  \author[KRISS]{K.~B.~Lee}
  \author[KRISS]{M.~K.~Lee}
  \author[Miyagi]{Y.~Fukuda}
  \author[Tokai1]{M.~Miyasaka}
  \author[Tokai1]{K.~Nishijima}
  \author[YNU1]{S.~Nakamura}
  
  \address[ICRR]{Kamioka Observatory, Institute for Cosmic Ray Research, the University of Tokyo, Higashi-Mozumi, Kamioka, Hida, Gifu, 506-1205, Japan}
  \address[IBS]{Center for Underground Physics, Institute for Basic Science, 70 Yuseong-daero 1689-gil, Yuseong-gu, Daejeon, 305-811, South Korea}
  \address[ISEE]{Institute for Space-Earth Environmental Research, Nagoya University, Nagoya, Aichi 464-8601, Japan}
  \address[Tokushima]{Institute of Socio-Arts and Sciences, The University of Tokushima, 1-1 Minamijosanjimacho Tokushima city, Tokushima, 770-8502, Japan}
  \address[IPMU]{Kavli Institute for the Physics and Mathematics of the Universe (WPI), the University of Tokyo, Kashiwa, Chiba, 277-8582, Japan}
  \address[KMI]{Kobayashi-Maskawa Institute for the Origin of Particles and the Universe, Nagoya University, Furo-cho, Chikusa-ku, Nagoya, Aichi, 464-8602, Japan}
  \address[Kobe]{Department of Physics, Kobe University, Kobe, Hyogo 657-8501, Japan}
  \address[KRISS]{Korea Research Institute of Standards and Science, Daejeon 305-340, South Korea}
  \address[Miyagi]{Department of Physics, Miyagi University of Education, Sendai, Miyagi 980-0845, Japan}
  \address[Tokai1]{Department of Physics, Tokai University, Hiratsuka, Kanagawa 259-1292, Japan}
  \address[YNU1]{Department of Physics, Faculty of Engineering, Yokohama National University, Yokohama, Kanagawa 240-8501, Japan}
  
  \fntext[Ogawanow]{Now at Department of Physics, College of Science and Technology, Nihon University, Kanda, Chiyoda-ku, Tokyo 101-8308, Japan.}
  \fntext[Yangnow]{Now at Center for Axion and Precision Physics Research, Institute for Basic Science, Daejeon 34051, South Korea.}
  \fntext[Tokushimanow]{Now at Department of Physics, Tokushima University, 2-1 Minami Josanjimacho Tokushima city, Tokushima, 770-8506, Japan}
  \fntext[Tohokunow]{Now at Research Center for Neutrino Science, Tohoku University, Sendai, Miyagi 980-8578, Japan.}

  \begin{abstract}
    XMASS-I is a single-phase liquid xenon detector
    whose purpose is direct detection of dark matter.
    To achieve the low background requirements necessary in the detector,
    a new model of photomultiplier tubes (PMTs), R10789,
    with a hexagonal window was developed based on the R8778 PMT
    used in the XMASS prototype detector.
    We screened the numerous component materials
    for their radioactivity.
    During development, the
    largest contributions to the reduction of radioactivity 
    came from 
    the stem and the dynode support.
    The glass stem was exchanged to the Kovar alloy one and 
    the ceramic
    support were changed to the quartz one.
    R10789 is the first model of Hamamatsu Photonics K. K.
    that adopted these materials for low background purposes and
    provided a
    groundbreaking step for further reductions of radioactivity in PMTs.
    Measurements with germanium detectors showed 
    1.2$\pm$0.3 mBq/PMT of $^{226}$Ra,
    less than 0.78 mBq/PMT
    of $^{228}$Ra,
    9.1$\pm$2.2 mBq/PMT of
    $^{40}$K, and 2.8$\pm$0.2 mBq/PMT
    of $^{60}$Co.
    In this paper, the radioactive details of the developed R10789 are described
    together with 
    our screening methods and the components of the PMT.
  \end{abstract}

\begin{keyword}
PMT \sep radioactivity  \sep HPGe \sep mass spectrometry 
\end{keyword}

\end{frontmatter}


\section{Introduction}
\label{sec:intro}
Direct detection of dark matter is one of the major scientific challenges in
modern astroparticle physics.
Based on the proposal 
\cite{org-xmass}, a single-phase liquid xenon (LXe) detector,
XMASS-I was constructed \cite{bibXMASSDetector}.
Since the dark matter signal is expected to be rare,
dark matter searches require low background detectors.
XMASS-I is designed
for dark matter and for other many rare event searches.
For a spin-independent WIMP-nucleon cross section, it is designed to search as
low as $10^{-45}$ cm$^{2}$ for a WIMP mass of 100 GeV/$c^{2}$.
To realize this sensitivity,
the background level
in the fiducial volume 
is required to be
~$10^{-4}$/day/kg/keV 
for deposited energies below 100 keV.

Background originating
from radioactive impurities in detector materials is
one of the most serious problems
in a low background experiment.
It is crucial to prepare
materials of sufficiently low radioactivity
before detector construction.

The dominant source of radioactivity 
in the XMASS prototype detector \cite{prototype}
was the Hamamatsu R8778
photomultiplier tubes (PMTs).
Though R8778 was a model developed for
the XMASS prototype detector 
and its radioactivity was much
lower than that of standard PMTs,
an order of magnitude reduction of radioactivity was necessary to meet
the background required for the XMASS-I detector. 
Therefore we developed  
a new model of low radioactivity PMT, R10789, based on R8778.
In order to lower the background, we screened
numerous candidate materials for their radioactivity.

In this paper
we describe
the development target in Section \ref{sec:target},
the screening methods in Section \ref{sec:method} and
the components of the PMT in Section \ref{sec:material}.
The radioactivity in
the final product of R10789 is presented in Section \ref{sec:result}.
The development summary is described in
Section \ref{sec:summary}, along with a 
discussion
outlining the largest component contributions to the radioactivity,
this is necessary to attain further reductions in
future developments.
The conclusion is written in Section \ref{sec:conclusion}.

\section{Target of the development}
\label{sec:target}
Firstly, we need
to define the activity level required for XMASS-I.
By using Monte Carlo simulations \cite{geant4} 
with the detector geometry at the design stage,
it is estimated that 
radioactivity one order of magnitude lower than the
R8778 PMT is required to reduce the
PMT contributions to a background
less than $2\times10^{-5}$/day/kg/keV
for deposited energies below 100 keV \cite{taup2007}.
Since the target of the XMASS-I detector's background is
$1 \times 10^{-4}$/day/kg/keV, this 20\% budget for the PMTs is a reasonable
requirement considering contributions from other sources of background.
Following this consideration,
we set
this as our target for the PMT development, outlined
in Table \ref{tab:r8778}.
The Kovar alloy used in both R8778 and R10789 
contains cobalt
therefore the target value of $^{60}$Co
is the same with that of the R8778 PMT.
$^{60}$Co is generated by ambient thermal neutrons inside the cobalt metal.
When R10789 was developed no other candidate material
except for Kovar was known 
to match the small thermal contraction coefficient of the
quartz window. 
We note that 
although the R8778 values listed in Table \ref{tab:r8778}
only refer to the PMT, the target values include both the PMT and 
a voltage divider circuit in this development 
as it is assumed negligible in comparison.
\begin{table*}[htb]
  \begin{center}
    \caption{The radioactivity of
      R8778 measured with HPGe detectors
      \cite{prototype} 
      and the target values
      for the development of the new PMT
      (1/10 of R8778, except for $^{60}$Co) in mBq/PMT.
      While the numbers for 
      R8778
      only reflect the PMT itself,
      the shown target numbers will be
      applied to the sum of the PMT and
      its voltage divider circuit.
      During the target estimation, we assumed the decay equilibrium is kept for
      the $^{238}$U chain and the $^{232}$Th chain.
    }
    \begin{tabular}{lcccc}
      \hline
      & The $^{238}$U chain 
      & $^{232}$Th chain & $^{40}$K & $^{60}$Co \\ 
      & ($^{238}$U, $^{226}$Ra) & ($^{232}$Th, $^{228}$Ra) & &\\
      \hline
      PMT(R8778) & 18$\pm$2 &  6.9$\pm$1.3  & 140$\pm$20 & 5.5$\pm$0.9 \\
      \hline
      \hline
      Target 
      & 1.8 & 0.69 & 14 & 5.5\\
      (PMT+circuit)  & & & &\\
      \hline
    \end{tabular}
    \label{tab:r8778}
  \end{center}
\end{table*}

\section{Material screening methods
  for radioactivity in the PMT}
\label{sec:method}

\subsection{High-purity germanium (HPGe) detector measurement}
To measure
the 
radioactivity of materials with sufficient sensitivity, we mainly used 
three low background HPGe detectors 
manufactured by CANBERRA. 
The detectors are installed at
the Kamioka observatory in the Kamioka Mine under Mt.
Ikenoyama, which provides 2700 m water
equivalent shielding from cosmic rays.
Two of them are co-axial p-type HPGe detectors with 100\% and 120\% relative detection
efficiency
\footnote{
  The efficiency is given relative to the efficiency of
  a three inch diameter 
  three inch thick NaI(Tl) for 1.332 MeV gamma from  $^{60}$Co source
  which is positioned 25 cm away from the detector.}
respectively, 
and the other is a co-axial n-type detector with 110\% relative
efficiency.
The sample chambers are cylindrical with a diameter of roughly 15 cm and a
height of 17 cm.
To suppress background caused by Rn gas,
these sample chambers are continuously flushed with
Rn-free air
in which the $^{222}$Rn concentration is less than a few tens
of mBq/m$^{3}$.
Lead (175 mm) and copper (50 mm)
shields cover the whole sample regions to
reduce background from the outside.

\begin{figure*}[htbp]
  \begin{center}
    \includegraphics[width=140mm]{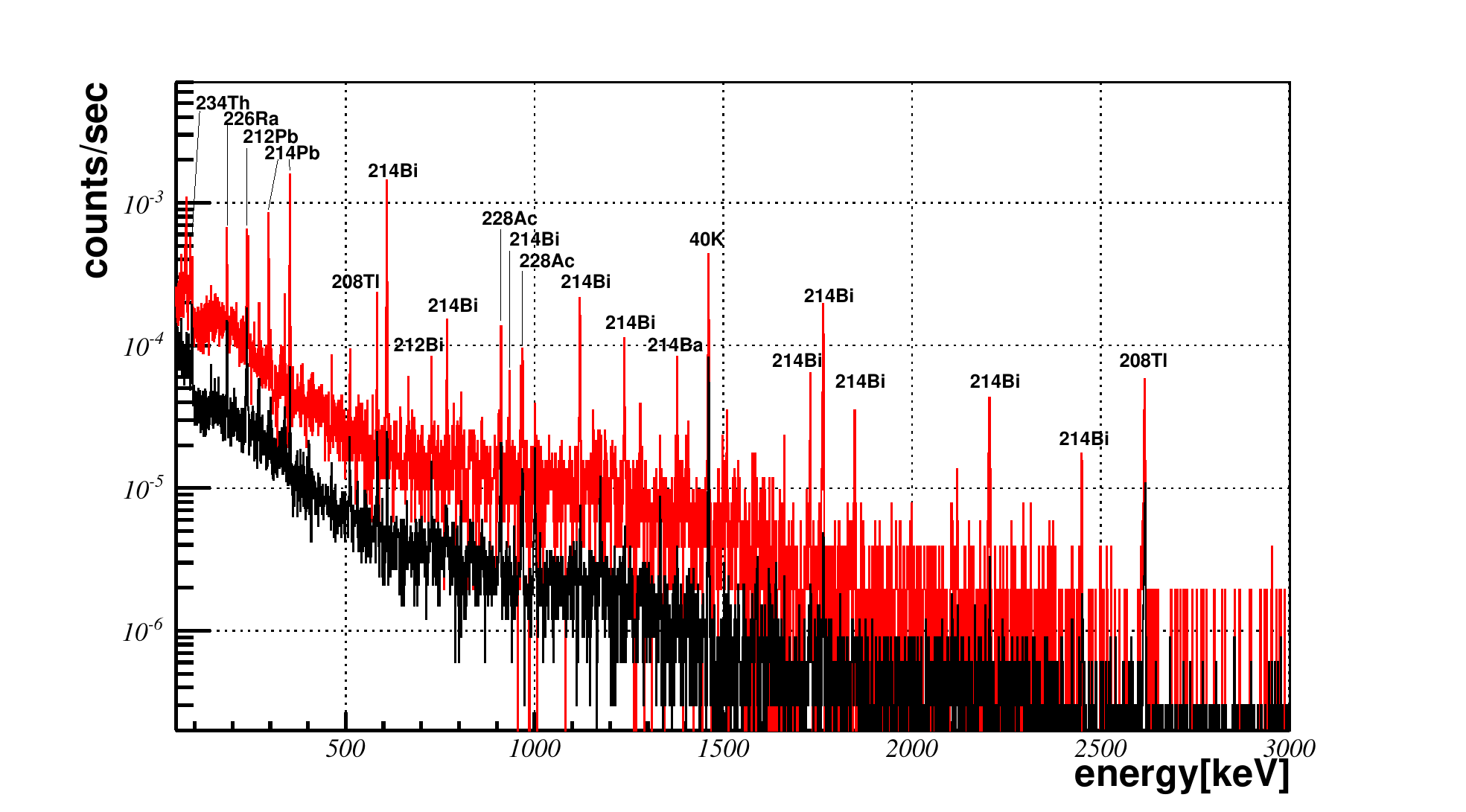}
  \end{center}
  \caption{
    Typical gamma-ray spectra
    of a glass beads sample and
    a background
    using one of our HPGe detectors are
    shown.
    The glass beads are largely radioactive
    as presented in Table \ref{tab:rpmts}, 
    1.61$\pm$0.02 mBq/PMT for $^{226}$Ra, 0.28$\pm$0.02 mBq/PMT for
    $^{228}$Ra
    and 3.3$\pm$0.2 mBq/PMT for $^{40}$K.
  }
  \label{fig:spectrum}
\end{figure*}

An example of measured gamma-ray
spectrum from one of our HPGe detectors
is shown in Figure
\ref{fig:spectrum} together with its
background spectrum.
The measured sample is glass beads
used as electrical insulation for the feedthroughs.
Many peaks from the $^{238}$U chain, the $^{232}$Th chain, and $^{40}$K can
be clearly identified. 
The energy region from 40 keV to 3 MeV
was examined to identify
radioisotopes
and to quantify their amount.
Typical sensitivities
of a few mBq/kg for the $^{238}$U and the $^{232}$Th chains are
obtained using a sample of a few kg with one week's worth of measurements.
Sensitivity for $^{60}$Co is
3 times better than
the $^{238}$U chain case and for
$^{40}$K is about 10 times worse 
than that of the $^{238}$U chain,
mainly due to the branching ratio to gamma rays
in its decay.

By comparison of actual data and Monte Carlo
simulation for three calibration sources, 
$^{57}$Co, $^{137}$Cs, and $^{60}$Co, 
we estimated the systematic
error of our HPGe measurements to be $+$30\%$-$10\%.
Since 
the systematic error is common for all measurements,
the results
in following sections are shown only with
their respective
statistical errors.

\subsection{Mass spectrometry for $^{238}$U and $^{232}$Th measurements}
\label{sec:massspec}
We also used two
methods from mass spectrometry, 
inductively coupled plasma mass spectrometry (ICPMS)
and glow discharge mass spectrometry (GDMS) for the
$^{238}$U and $^{232}$Th measurements.

Long-lived isotope often disrupt a decay chain's
equilibrium , as in the $^{238}$U and the $^{232}$Th chains.
For the $^{238}$U chain we measured
$^{238}$U for the top part and $^{226}$Ra for
the middle part of the $^{238}$U chain independently.
For the $^{232}$Th chain $^{232}$Th and
$^{228}$Ra for the part of after $^{232}$Th were measured.
Since the branching ratios to high energy gamma-rays,
which are suitable for HPGe measurements, are small,
the HPGe's sensitivities to
the nuclear species before $^{226}$Ra for U chain and
to $^{232}$Th
are very low.
From these nuclear species, only low energy gamma-rays, beta-rays,
and alpha rays
are emitted, and due to their short interaction
length, the contribution to the detector background is limited to
radioisotopes that are very close
to the LXe inside XMASS-I
\footnote{
  It is estimated that in XMASS-I,
    the background originated from neutrons as fissions fragments of $^{238}$U
    is much smaller and can be negligible compared to the one from the decay chain. 
}
.
For this reason, $^{238}$U and $^{232}$Th measurements by mass spectrometry were
conducted only for the PMT components at the inner side of the detector,
not for the voltage divider circuit.
\section{Components of the PMT, R10789, and the voltage divider circuit}
\label{sec:material}
In this section, we
discuss the various component materials used in the newly
developed R10789
PMT 
and its voltage divider circuit.

\subsection{The PMT(R10789)}
The PMTs are divided into ten components:
dynodes, electrode, two different
body cylinders, stem, glass beads, lead wires, quartz, 
sealing, and getter.
Measurements were conducted
on the sample for each component.

\begin{figure}[htbp]
  \begin{center}
    \includegraphics[width=60mm]{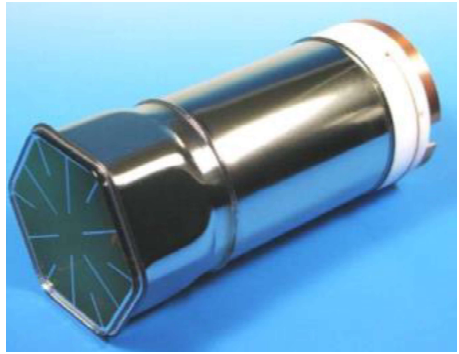}
  \end{center}
  \caption{The R10789 PMT with the voltage divider circuit.}
  \label{fig:r10789}
\end{figure}

\begin{figure}[htbp]
  \begin{center}
    \includegraphics[width=60mm]{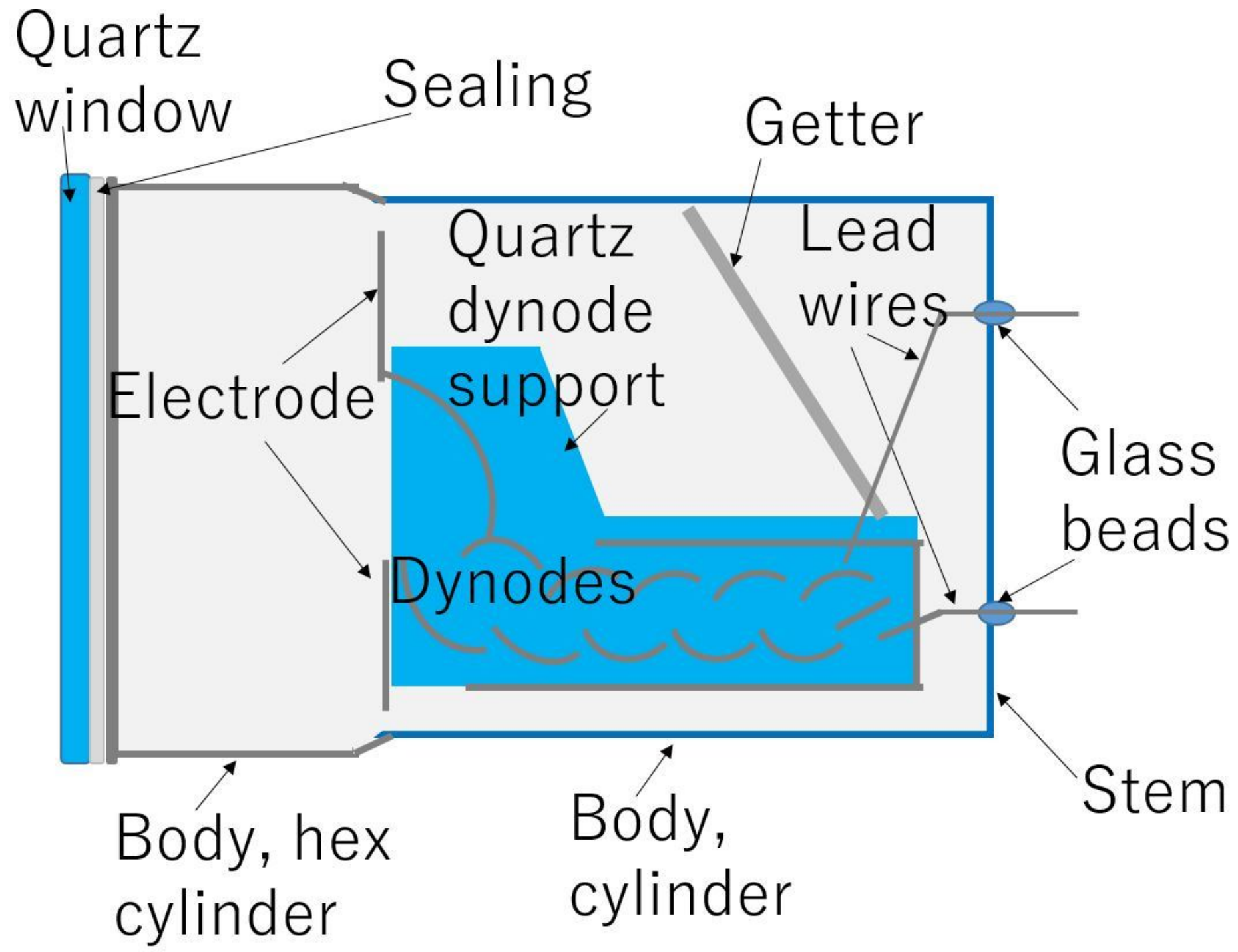}
  \end{center}
  \caption{The schematic view of PMT R10789. The components
    as explained in
    Section \ref{sec:material} are shown.} 
  \label{fig:pmtf}
\end{figure}

These components are listed in Table \ref{tab:pmts},
together with the weights
of the measurement samples and the respective amount used to produce 
one PMT.
The weights for the HPGe measurements were a factor
4-100 larger than
what is used for one PMT.
Figure \ref{fig:r10789} shows the R10789 PMT with the voltage
divider circuit, and 
Figure \ref{fig:pmtf} shows a schematic view of R10789.

There are two body components
used in one R10789 PMT,
both of which are made from Kovar alloy.
One is a hexagonal-shape tube
matching the shape of the quartz entrance window, 
while the rest of the body is cylindrical
to which the stem is attached.
The materials of the dynode support and the stem were exchanged from the ceramic
and the glass in R8778 to the quartz and the Kovar alloy in R10789,
respectively.
The ceramic and the glass were the largest sources
of radioactivity in the R8778 PMT, by reducing the amounts of the ceramic
and glass in the new R10789 PMT
as much as possible, a large reduction of
radioactivity was achieved.

\subsection{The voltage divider circuit}
Items in the voltage divider circuit, 
which include
circuit parts such as resistors, capacitors and a 
circuit board are also presented in
Table \ref{tab:pmts}. 
Two cables, one for signal and
the other for the
high voltage (HV)
supply, are also included in this table.
Since most of the parts are small, it was possible
to measure large quantities of the parts 
used for a few tens to hundreds of the PMTs in one measurement. This led to
high precision measurements.

The cables are polytetrafluoroethylene (PTFE) insulated.
The signal coaxial cable is compatible with RG196, and
both the
inner dielectric insulator and the outer
sheath are made from PTFE.
The HV cable is an AWG 22 copper conducting wire covered with PTFE.
The actual cable length in XMASS-I is about 10 m,
but 
only the effects from ~1 m near the detector surface contribute,
we therefore listed the radioactivity for 1 m of the cable in Table \ref{tab:pmts}.

\begin{table*}[htb]
  \begin{center}
    \caption{List of the
    developed R10789 parts. The
    weights for HPGe measurements and
    the weight of each component per PMT
    are presented together.
    Total weight of the parts corresponding to the one R10789 is 162.8 g.}
  \begin{tabular}{lcc}
    \hline
    Samples & Weight for & Weight per\\ 
    &  measurement (g) & PMT (g)\\
    \hline
    \hline
    PMT(R10789) & & \\
    \hline
    Dynodes (stainless steel) & 159 & 16.2\\
    Electrode (stainless steel) & 89.3  & 8.22\\
    Body, hex cylinder (Kovar alloy) & 144  & 34.1\\
    Body, cylinder (Kovar alloy) & 188 & 49.7\\
    Stem (Kovar alloy) & 289& 26.7\\
    Glass beads & 164 & 1.81\\
    Lead wires (Nickel) &181 & 1.78\\
    Quartz (Window \& dynode support) & 114 & 23.9\\
    Sealing (Aluminum) & 10.0 & 0.336 \\
    Getter  & 1.0 & 0.07 \\
    \hline
    \hline
    Voltage divider circuit and Cables &  &\\ 
    \hline
    Resistors KTR10EZPF & 20.4 (4300 pieces) & 0.0810 (17 pieces)\\ 
    Decoupling capacitors ECWU-JC9 & 3.56 (80 pieces) & 
    0.178 (4 pieces)\\
    Coupling capacitors C4520X7R3D & 12.7 (200 pieces)  & 0.127 (2 pieces)\\
    Circuit board (polyimide) & 432  & 0.99\\
    Solder Sr34 LFM48 (Sn, Ag, Cu) & 25.8 & 0.362 \\
    Socket connectors (Brass) & 16.3 (300 pieces) & 1.09 (20 pieces) \\
    Screw m2 (Stainless steel) & 11.2 (100 pieces) & 0.448 (4 pieces) \\
    Terminal for HV cables (Brass) & 118 (400 pieces) & 0.591 (2 pieces) \\
    Signal connectors MCX (Brass) & 181 (100 pieces) &
    1.81
    (1 piece) \\
    PTFE fillers & 2540  & 66.3 \\
    Signal cable & 500 (50 m) & 10.0 (1 m) \\
    HV cable & 670 (100 m) & 6.70 (1 m) \\
    \hline
    \hline
  \end{tabular}
  \label{tab:pmts}
  \end{center}
\end{table*}

\section{Radioactivity for the developed R10789 PMT}
\label{sec:result}
The total radioactivity of
the developed R10789
is summarized in this section,
the unit of activity is
mBq/PMT.

\subsection{The results from measurement of each part}
\label{sec:pmtresults}
Tables
\ref{tab:rpmts} and \ref{tab:rbases} show the results of the R10789
PMT and the voltage divider circuit
using the HPGe detectors. 
In case the center value 
does not exceed twice
the statistical error, or is negative, 
we regarded the result as zero
consistent and set an upper limit
calculated as
$\max\{0, {\mbox{the central value}}\}+1.28\times \mbox{statistical error}$.  
The total radioactivity for the PMT components
and the voltage divider circuit components are 
shown in Table \ref{tab:rsums} along
with their target values introduced above. 
In the calculation of the total radioactivity 
the central values of each component are summed up,
each value's sign is included in the sum.

While the values for 
$^{40}$K and $^{60}$Co are 
significantly smaller than their targets,
$^{226}$Ra and $^{228}$Ra 
are larger by a factor of 1.3 and 2, respectively. \\
For $^{226}$Ra, 
the largest contribution comes from the glass beads, which accounts
for about 70\% of the total radioactivity 2.3$\pm$0.3 mBq/PMT.  
Other large contributions come from the coupling capacitors and the signal cable,
about 10\% each.
These three components dominate and account for about 90\% of the total.\\
As for the $^{228}$Ra,
the largest detected value is again
from the glass beads, about 20\% of the total
radioactivity, 1.6$\pm$0.3 mBq/PMT.
The getter, the coupling capacitors, the signal connectors and the signal cables
also make non-negligible contribution
of about 0.1 mBq/PMT
\footnote{Sum of the
  detected values is only a half of the total sum,
  the
  remaining half comes from many large center values in
  the
  upper limits due to their statistics error.
  While each value are zero consistent within error, the sum of them
  exceeds
  twice of
  error.}.\\
The glass beads and the resistors are the only
samples from which significant $^{40}$K are detected.
The total $^{60}$Co activity is
only  ~60\% of the target.
The main contribution comes
from the Kovar alloy in 
the two PMT body parts and the stem.
Another 
$^{60}$Co contribution comes from
the dynodes, however it 
is quite small compared to the body.

\begin{landscape}
\begin{table*}[htb]
  \begin{center}
    \caption{Results of 
    measurements
    for the developed R10789 parts measured in mBq/PMT for
    $^{226}$Ra, $^{228}$Ra, $^{40}$K
     and $^{60}$Co with HPGe, and 
     for $^{238}$U and $^{232}$Th
     with HPGe, GDMS and ICPMS.
    Measuring methods for $^{238}$U and $^{232}$Th are also listed.
    Details of the calculation of upper limit
      and summation are explained in the text.
  }
  \begin{tabular}{lccccccc}
    \hline
    Samples & $^{226}$Ra  & $^{228}$Ra & $^{40}$K & $^{60}$Co & $^{238}$U &
    $^{232}$Th
    & Method for\\ 
    & & & & & & & $^{238}$U \& $^{232}$Th\\
    \hline
    \hline
    PMT(R10789)& & & & & &\\
    \hline
    Dynodes     & \(<\)0.10 & \(<\)0.27 & \(<\)0.95 &0.11$\pm$0.04&\(<\)2.8\,$\cdot\,10^{-2}$ & 0.15 & GDMS\\
    Electrode   & \(<\)0.20 & \(<\)0.17 & \(<\)0.63 & \(<\)8.8\,$\cdot\,10^{-2}$& \(<\)1.0\,$\cdot\,10^{-2}$ & \(<\)3.3\,$\cdot\,10^{-3}$ & GDMS\\
    Body, hex part   & \(<\)0.16 & \(<\)0.35 & \(<\)1.6 & 1.0$\pm$0.1& \(<\)4.2\,$\cdot\,10^{-2}$ & \(<\)1.4\,$\cdot\,10^{-2}$ & GDMS\\
    Body, cylinder part & \(<\)0.18 & \(<\)0.33 & \(<\)1.6 & 1.14$\pm$0.08& \(<\)6.2\,$\cdot\,10^{-2}$ & \(<\)2.0\,$\cdot\,10^{-2}$ & GDMS\\
    Stem             & \(<\)0.14 & \(<\)0.24 & \(<\)1.2 & 1.08$\pm$0.07& \(<\)3.3\,$\cdot\,10^{-2}$ & \(<\)1.1\,$\cdot\,10^{-3}$ & GDMS\\
    Glass beads                       & 1.61$\pm$0.02 & 0.28$\pm$0.02 & 3.3$\pm$0.2 & \(<\)6.7\,$\cdot\,10^{-3}$& 0.43$\pm$0.16 & &  HPGe\\
    Lead wires        & \(<\)9.4\,$\cdot\,10^{-3}$ & \(<\)2.3\,$\cdot\,10^{-2}$ & \(<\)8.2\,$\cdot\,10^{-2}$ & \(<\)4.0\,$\cdot\,10^{-3}$& \(<\)2.2\,$\cdot\,10^{-3}$ & \(<\)7.2\,$\cdot\,10^{-4}$ &GDMS\\
    Quartz  & \(<\)0.18 & \(<\)0.19 & \(<\)2.9 & \(<\)7.5\,$\cdot\,10^{-2}$&\(<\)6.0\,$\cdot\,10^{-3}$ & \(<\)1.9\,$\cdot\,10^{-3}$ & ICPMS\\
    Sealing          & \(<\)1.2\,$\cdot\,10^{-2}$ & (3.5$\pm$0.9)\,$\cdot\,10^{-2}$ & \(<\)0.238 & \(<\)6.0\,$\cdot\,10^{-3}$& 1.3 & 2.5\,$\cdot\,10^{-2}$ & ICPMS\\
    Getter                           & \(<\)6.1\,$\cdot\,10^{-2}$ & 0.10$\pm$0.04 & \(<\)0.53 & \(<\)2.2\,$\cdot\,10^{-2}$ & &\\
    \hline
    Sum of the PMT & & & &\\
    (R10789) parts         & 1.6$\pm$0.3 & 1.1$\pm$0.3 & \(<\)3.2 & 3.4$\pm$0.2 \\
    \hline
  \end{tabular}
  \label{tab:rpmts}
  \end{center}
\end{table*}

\begin{table*}[htb]
  \begin{center}
  \caption{Results of HPGe measurements
    for the developed voltage divider circuit parts measured in mBq/PMT for
    $^{226}$Ra, $^{228}$Ra, $^{40}$K and $^{60}$Co.
  }
  \begin{tabular}{lcccc}
    \hline
    Samples & $^{226}$Ra  & $^{228}$Ra & $^{40}$K & $^{60}$Co \\ 
    \hline
    \hline
    Voltage divider circuit and Cables& & & & \\
    \hline
    Resistors KTR10EZPF            & (3.3$\pm$0.4)\,$\cdot\,10^{-2}$ & (3.5$\pm$0.4)\,$\cdot\,10^{-2}$ & 0.34$\pm$0.05 & \(<\)2.0\,$\cdot\,10^{-3}$  \\ 
    Decoupling capacitors ECWU-JC9 & \(<\)2.4\,$\cdot\,10^{-2}$ &\(<\)3.0\,$\cdot\,10^{-2}$ &\(<\)0.41 &\(<\)1.2\,$\cdot\,10^{-2}$ \\
    Coupling capacitors C4520X7R3D & 0.22$\pm$0.03 & (8.8$\pm$1.8)\,$\cdot\,10^{-2}$ & \(<\)0.14 & \(<\)6.7\,$\cdot\,10^{-3}$\\
    Circuit board   & (2.3$\pm$0.4)\,$\cdot\,10^{-2}$ & \(<\)8.2\,$\cdot\,10^{-3}$ & \(<\)4.0\,$\cdot\,10^{-2}$ & \(<\)1.7\,$\cdot\,10^{-3}$\\
    Solder Sr34 LFM48 & \(<\)2.6\,$\cdot\,10^{-2}$ & \(<\)2.0\,$\cdot\,10^{-2}$ & \(<\)0.44 & \(<\)5.8\,$\cdot\,10^{-3}$ \\
    Socket connectors  & \(<\)5.0\,$\cdot\,10^{-2}$ & \(<\)8.8\,$\cdot\,10^{-2}$ & \(<\)0.58 & \(<\)3.9\,$\cdot\,10^{-2}$\\
    Screw m2   & \(<\)3.1\,$\cdot\,10^{-2}$ & (4.9$\pm$1.4)\,$\cdot\,10^{-2}$ & \(<\)0.19 & \(<\)7.5\,$\cdot\,10^{-3}$\\
    Terminal for HV cables & (7.4$\pm$3.6)\,$\cdot\,10^{-3}$ & \(<\)1.3\,$\cdot\,10^{-2}$ & \(<\)5.0\,$\cdot\,10^{-2}$ & \(<\)1.9\,$\cdot\,10^{-3}$\\
    Signal connectors MCX  & \(<\)2.8\,$\cdot\,10^{-2}$ & (8.4$\pm$2.1)\,$\cdot\,10^{-2}$ & \(<\)0.18 & \(<\)1.2\,$\cdot\,10^{-2}$ \\
    PTFE fillers                   & \(<\)0.22 & \(<\)0.19 & \(<\)1.3 & \(<\)3.4\,$\cdot\,10^{-2}$ \\
    Signal cable               & 0.31$\pm$0.04 & 0.15$\pm$0.03 & \(<\)0.37 & \(<\)2.0\,$\cdot\,10^{-2}$ \\
    HV cable                    & \(<\)5.0\,$\cdot\,10^{-2}$     & \(<\)5.6\,$\cdot\,10^{-2}$     & \(<\)1.00  & \(<\)2.9\,$\cdot\,10^{-2}$ \\
    \hline
    Sum of the voltage divider circuit parts & 0.73$\pm$0.11 & 0.55$\pm$0.11 &  \(<\)1.6 &  \(<\)5.2\,$\cdot\,10^{-2}$\\
    \hline
  \end{tabular}
  \label{tab:rbases}
  \end{center}
\end{table*}
\end{landscape}

\begin{table*}[htb]
  \begin{center}
    \caption{Sum of the HPGe results for the PMT(R10789) and the voltage divider circuit.
    The target numbers are listed also. The unit is mBq/PMT.
  }
  \begin{tabular}{lcccc}
    \hline
     & $^{226}$Ra  & $^{228}$Ra & $^{40}$K & $^{60}$Co \\ 
    \hline
    \hline
    Sum of the PMT(R10789) & 2.3$\pm$0.3 & 1.6$\pm$0.3 & \(<\)3.5 & 3.3$\pm$0.2 \\
    and the circuit & & &\\
    \hline
    \hline
    Target (PMT+circuit)& 1.8 & 0.69 & 14 & 5.5\\
    \hline
  \end{tabular}
  \label{tab:rsums}
  \end{center}
\end{table*}

\subsection{Results for the assembled PMT}
We also carried out measurements
of assembled PMTs from all of the parts
to check against contamination during
the assembly processes. 
Three assembled PMTs were measured by the HPGe,
and the results are shown in Table \ref{tab:rpmt}.
A larger amount of $^{40}$K was detected in the assembled PMTs. 
This is presumably because 
the chemicals used for manufacturing the photo-cathode 
contain potassium, these chemicals were not
included in the components measurements.
Except for $^{40}$K,
the summed component and assembled PMT
measurements are consistent with each other,
and thus we concluded
that there was no significant
contamination in the assembly processes.

\begin{table*}[htb]
  \begin{center}
  \caption{Results of the three assembled R10789 PMTs
    with the HPGe detector.
    The sum of the radioactivity in the PMT components in
    Table \ref{tab:pmts} are also presented for the reference. 
    The unit is mBq/PMT.
   }
  \begin{tabular}{lcccc}
    \hline
    Samples & $^{226}$Ra  & $^{228}$Ra & $^{40}$K & $^{60}$Co \\ 
    \hline
    Assembled PMT (R10789)      & 1.2$\pm$0.3 & \(<\)0.78 & 9.1$\pm$2.2 & 2.8$\pm$0.2\\
    \hline
    Sum of the PMT parts (R10789)   & 1.6$\pm$0.3 & 1.1$\pm$0.3 & \(<\)3.2 & 3.4$\pm$0.2 \\
    \hline
  \end{tabular}
  \label{tab:rpmt}
  \end{center}
\end{table*}

\subsection{Results from mass spectrometry}
Table \ref{tab:rpmts} shows the results for $^{238}$U and $^{232}$Th.
As for $^{238}$U, 
except for the glass beads and the aluminum sealing,
the measured  upper limits of the
top part of the $^{238}$U chain are much smaller 
and consistent 
with the results of $^{226}$Ra.
The differences between the two measurements of
the glass beads and the aluminum sealing 
demonstrate
that within these components, the
decay equilibrium of the $^{238}$U chain is 
broken. 
The sum of $^{238}$U in the glass beads and the aluminum sealing
is large at about 1.5 mBq/PMT,
and thus the contributions from the other components
can be ignored.
The target for $^{238}$U is the same as $^{226}$Ra,
1.8 mBq/PMT.\footnote{We assumed the secure equilibrium in our
  Monte Carlo simulation.}\\
All results of $^{232}$Th are smaller and consistent with the results of $^{228}$Ra.
As in the case of $^{238}$U, the target for $^{232}$Th is the same as $^{228}$Ra.\\
Since both sums of $^{238}$U and $^{232}$Th are smaller than $^{226}$Ra and $^{228}$Ra,
it is enough to compare only the results of $^{226}$Ra and $^{228}$Ra to each target value,
and will be discussed in the next session.

\section{Discussion}
\label{sec:summary}
\subsection{Comparison to the target numbers}
\label{sec:compare}
As a result of the development,
a significant reduction of radioactivity 
compared to R8778 was achieved.
The activities achieved for $^{40}$K and $^{60}$Co
are much smaller than their target values.
The $^{40}$K measurement of 9.1$\pm$2.2 mBq/PMT for 
the assembled PMT shown in Table \ref{tab:rpmt}, 
is larger than the sum of all its parts,
however, is still smaller than
its target.
As already mentioned, the
larger amount of $^{40}$K observed in the assembled PMT 
can be
understood
as
a likely extra
contribution
coming
from the photo-cathode material,
which is introduced inside the PMT
in the assembly process.
Since it is very difficult to estimate the amount of potassium
left inside of the PMT after the vapor deposition,
a further investigation
was not undertaken.
\\
The final
activities for
$^{226}$Ra and $^{228}$Ra  in the newly developed PMT and
its voltage divider circuit 
exceed the target values, especially $^{228}$Ra
which is a factor
of two larger.
Nevertheless, we should recall that the target numbers were
set with a
large tolerance of
at least a factor of
five, and therefore the activities are
allowable for the background target in XMASS-I,
$10^{-4}$/day/kg/keV.
These newly developed PMTs are used in the XMASS-I detector,
by which many dark matter and rare event searches have been carried out
\cite{doublee, modulation, kkaxion, swimp, inela, axion, lwimp}.
The observed amount of background caused by these PMTs
in the XMASS-I detector
is discussed in \cite{FVpaper}.

\subsection{Components which have largest contributions}
\label{sec:largest}
It is valuable to summarize key improvements and significance of
  this development.
The largest contributions to the reduction came from  
two items, the stem and the dynode support.
In R8778, the stem
was made of glass and the dynode support from ceramic.
In the new PMT, R10789, Kovar alloy and quartz are used respectively.
By exchanging
the glass stem for one made from Kovar alloy,
$^{226}$Ra was
reduced from 2.3$\pm$0.1 mBq/PMT to \(<\)0.14 mBq/PMT,
and 
$^{228}$Ra from 2.3$\pm$0.1 mBq/PMT to \(<\)0.24 mBq/PMT.
By using quartz instead of ceramic for the dynode support,
$^{226}$Ra was reduced from 7.9$\pm$0.2 mBq/PMT to
\(<\)6.6\,$\cdot\,10^{-2}$ mBq/PMT and 
$^{228}$Ra from 3.1$\pm$0.2 mBq/PMT
to \(<\)7.0\,$\cdot\,10^{-2}$ mBq/PMT.
The R10789 is the first PMT of Hamamatsu Photonics K. K.
adopting the Kovar alloy stem and
the quartz dynode support 
for low background purposes.
This knowledge and
these materials were adopted in the development of R11410
\cite{xenon100,lux,pandax,xenon1t}
and R13111 \cite{R13111} with improved background.
This clearly indicates the development of R10789 is a groundbreaking step for
further improvement of radioactivity in PMTs.

\subsection{For further reductions}
\label{sec:future}
To reduce the radioactivity even further in much lower
background experiments, 
it is necessary to deal with the 
high radioactive components.
The glass beads include large amounts
of $^{226}$Ra, $^{228}$Ra and $^{40}$K, $^{226}$Ra
accounts for 70\% of the total. 
Some parts of the voltage divider circuit,
such as the coupling capacitors,  and the signal cable
also have a large amount of $^{226}$Ra and $^{228}$Ra.
In the capacitors and the signal cables, ceramic and PTFE are
used respectively as raw materials.
These two materials and glass mentioned above
generally contain non-negligible amounts
of radioactive impurities.
Therefore, to use these materials,
it is quite essential to find materials 
with lesser radio impurities and to reduce their usage as much as possible.
\\
As for $^{40}$K, a larger amount was found
in the assembled PMTs
than was expected from the sum of its parts,
this is speculated to come
from the photocathode.
Chemicals used while treating
the photocathode contain
potassium, and therefore it is not easy to suppress this
$^{40}$K contamination
but to control it would be essential for next generation ultra-low-background PMTs.  
To reduce $^{60}$Co, Kovar alloy parts (body cylinders and stem) must be
replaced with  a new material
whose thermal contraction coefficient matches that of 
quartz and does not contain cobalt.
The Co-free alloy which is used for example
  by the R11410 PMT \cite{xenon1t} could be used.
\\
In our screening, there were many items where 
only upper limits were set.
These could contribute significantly in 
future PMT 
developments, especially $^{228}$Ra and $^{40}$K. 
Higher sensitivity measurements
are strongly demanded for further reductions.

\section{Conclusion}
\label{sec:conclusion}
We succeeded in developing the new low
background PMT, R10789, which satisfies  
the requirements of the XMASS-I detector.
It achieved large reduction
factor of 8, 4, 1.7 and more than 10 from PMT R8778
for  $^{226}$Ra, $^{228}$Ra, $^{60}$Co, and $^{40}$K, respectively.
These facts prove that it can be 
utilized in various low background experiments.
The largest contributions to the reduction were
obtained 
by exchanging two items,
the stem glass with Kovar alloy
and the ceramic dynode support with quartz.
This is the first model of Hamamatsu Photonics K. K. PMTs 
that adopted these materials for low background purposes
and provided
a groundbreaking step for further improvements of
radioactivity in PMTs.

\section*{Acknowledgments}
We thank Hamamatsu Photonics K. K. for the cooperation in producing the low RI PMTs.
We gratefully acknowledge the cooperation of Kamioka Mining and Smelting Company.
This work was supported by the Japanese Ministry of Education, Culture, Sports, Science and
Technology, Grant-in-Aid for Scientific Research, ICRR Joint-Usage, JSPS KAKENHI Grant Number,
19GS0204 and 26104004, 
and partially
by the National Research Foundation of Korea Grant (NRF-2011-220-C00006) 
and Institute for Basic Science (IBS-R017-G1-2018-a00).

\end{document}